\documentclass[12pt]{iopart}

\usepackage{iopams}
\usepackage{setstack}
\usepackage{graphicx}
\usepackage{amsfonts}
\usepackage{textcomp}

\newcommand{\BEA}{\begin{eqnarray}}
\newcommand{\EEA}{\end{eqnarray}}

\usepackage{mathrsfs}
\usepackage{color}
\usepackage{xcolor}

\newcommand{\ket}[1]{\left|#1\right>} 
\newcommand{\bra}[1]{\left<#1\right|} 

\newcommand{\brr}{\boldsymbol{r}}

\newcommand{\be}{\boldsymbol{e}}
\newcommand{\bhe}{\bi{e}}
\newcommand{\bS}{\boldsymbol{S}}
\newcommand{\bE}{\boldsymbol{E}}

\newcommand{\ce}{\mathrel{\mathop:}=}

\begin{document}

\title{Classical entanglement in polarization metrology}
\author{Falk T\"{o}ppel$^{1,2,3}$, Andrea Aiello$^{1,2}$, Christoph Marquardt$^{1,2}$, Elisabeth Giacobino$^{1,4}$ and Gerd Leuchs$^{1,2}$}
\ead{${}^*$falk.toeppel@mpl.mpg.de} 
\address{$^1$ Max Planck Institute for the Science of Light, G\"{u}nther-Scharowsky-Stra{\ss}e 1/Bldg. 24, 91058 Erlangen, Germany}
\address{$^2$ Institute for Optics, Information and Photonics, Universit\"{a}t Erlangen-N\"{u}rnberg, Staudtstra{\ss}e 7/B2, 91058 Erlangen, Germany}
\address{$^3$ Erlangen Graduate School in Advanced Optical Technologies (SAOT), Paul-Gordan-Stra{\ss}e 6, 91052 Erlangen, Germany}
\address{$^4$ Laboratoire Kastler Brossel, Universit\'e Pierre et Marie Curie, Ecole Normale Sup\'erieure, CNRS, 4 place Jussieu, 75252 Paris Cedex 05, France}
\date{\today}
\begin{abstract}
Quantum approaches relying on entangled photons have been recently proposed to increase the efficiency of optical measurements. We demonstrate here that, surprisingly, the use of classical light with entangled degrees of freedom can also bring outstanding advantages over conventional measurements in polarization metrology. Specifically, we show that radially polarized beams of  light allow to perform real-time single-shot Mueller matrix polarimetry.
Our results also indicate that quantum optical procedures requiring entanglement without nonlocality can be actually achieved in classical optics regime.

\end{abstract}
\pacs{03.65.Ud, 41.20.-q, 42.25.Ja}

\maketitle

%
\section{Introduction}
In the last years, quantum information theory taught us that the use of entangled photons offers the unique advantage over classical light of providing more information in metrology applications, imaging and, more generally, in optical measurements \cite{Massar,Gisin,Abouraddy1,Aiello2004,Legre,Brunner}.
However, entanglement is not necessarily a signature of the quantum mechanical nature of a system.
Indeed, one can distinguish between
two types of entanglement: \textit{a}) entanglement between spatially separated systems (\emph{inter}-system entanglement) and \textit{b}) entanglement between different degrees of freedom (DoFs) of a single system (\emph{intra}-system entanglement) \cite{Gisin1991,Spreeuw}.
Inter-system entanglement, or ``nonlocal entanglement'',  can occur only in bona fide quantum systems and may yield to nonlocal statistical correlations. Conversely, intra-system entanglement, or ``local entanglement'',  may also appear in classical systems and can not generate nonlocal correlations  \cite{Gisin2005}.
As an example, photon pairs from atomic cascades \cite{Aspect} show nonlocal entanglement. On the opposite, local entanglement can be found, e.g., between spatial and spin DoFs in single neutrons \cite{Hasegawa}. 
In classical optics, local entanglement between polarization and spatial DoFs of the same beam, has been lately demonstrated in radially and azimuthally polarized beams of light \cite{Khoury2007,Luis,HolleczekOL,Khoury,Karimi2010,EberlyPlag}. Hereafter, we shall denote the occurrence of local entanglement in classical systems with ``classical entanglement'' \cite{Spreeuw}\footnote{In the literature this phenomenon is also refereed to as ``structural inseparability'' \cite{Holleczek,Gabriel2011} and ``nonquantum entanglement'' \cite{Simon}.}.

Recently, Khoury and coworkers have suggested that quantum computing
tasks requiring entanglement but not nonlocality can be efficiently accomplished in the classical optical regime  \cite{Khoury}. However,  when and how classical entanglement can be exploited in lieu of nonlocal entanglement to improve techniques of optical measurements, still remain largely open questions  \cite{Walborn,Gianni,Ghose}.

In this work, we address some of these central issues by proposing to use classical entanglement in radially polarized beams of light for highly efficient Mueller matrix polarimetry \cite{Mueller,Damask,Jeune,Aiello2006}.
The underlying idea is simple: In a conventional Mueller matrix measurement setting, an either transmissive or scattering material sample (the object) is illuminated with a light beam (the probe) prepared in, at least, \emph{four} different polarization states in a temporal sequence.
From the analysis of the polarization of the light transmitted or scattered, the optical properties of the object
can be inferred.
In the alternative setting we propose here,
the object is probed only once with \emph{one} light beam of radial polarization, as opposed to four differently polarized beams. Then, the light transmitted or reflected by the object is analyzed \emph{both} in polarization and in spatial DoFs by means of suitable polarization and spatial mode selectors.
In our setting, the polarization DoFs of the beam are used to actually probe the object and the spatial DoFs
are used to postselect the polarization state of the light. This scheme outperforms conventional ones because
the radially polarized beam carries all polarizations at once in a classically entangled state, thus providing for a sort of
``polarization parallelism''.

Hence, while with conventional Mueller matrix measurements it is necessary to probe the object fourfold, by using radially polarized light one can obtain the same amount of information by probing the sample only once. Therefore, for all practical applications where the optical properties of the sample changes rapidly with time, our method presents an outstanding advantage over conventional ones. We believe that this result can provide significant improvements in polarization metrology applications \cite{Mueller,Damask,Nuclear,Stellar,Geo,Biomedical,Sensing,Tissue}.
Although our conclusions here are strictly valid only for optical elements that do not alter significantly the spatial structure of the probe beam, this is not a serious restriction.
For example, all optical elements routinely used on an optical bench like single-mode fibers, retardation plates, birefringent prisms, optical rotators, etc.,   fall in this category.
Moreover, our method can be also straightforwardly implemented by using less constrained entangled binary DoFs as, e.g., the polarization and spatial-parity modes considered by Saleh and coworkers \cite{Gianni,Abouraddy}. Another requirement for the validity of our scheme, is that the polarization properties of the sample must be homogeneous over the beam cross section. Considering that an ordinary radially polarized beam of light may be prepared with a waist of the order of hundreds of \textmu m, such a requirement does not represent an actual limitation.
%
%
\section{Jones vectors, entanglement and radially polarized beams}
Consider a  monochromatic beam of light of angular frequency $\omega$, propagating along the  $z$-axis of a Cartesian reference frame $(x,y,z)$ and polarized in the $(x,y)$ plane. In the paraxial approximation \cite{Siegman}, the electric field can be written as $\bi{\mathcal{E}}(\brr,t) = \mathrm{Re}\left[\bE(\brr)\exp(-i \omega t) \right]$, where
\BEA
\label{PW_field}
\bE(\brr)= \left( A_0 \, \bhe_x +A_1 \bhe_y \right) \! \psi(\brr),
\EEA
with $\psi(\brr)$ denoting the spatial mode of the beam, $\bhe_x$, $\bhe_y$ and $\bhe_z$ being unit vectors in the $x$, $y$ and $z$ directions, respectively.
In the expression above,   $\brr = x \bhe_x +y \bhe_y + z \bhe_z$ stands for the position vector,
and the two complex numbers $A_0$ and $A_1$ represent the amplitudes of the electric field along the $x$ and $y$ axis, respectively.
A convenient vector notation for fields of the form (\ref{PW_field}) was introduced by R. C. Jones in the 1940s \cite{Damask,Jones}:
\BEA
\label{PW_field_Jones}
\bE(\brr)= \left[\begin{array}{c}
 \!   \! A_0 \!  \! \\
  \! \!  A_1 \!  \!
\end{array}\right] \! \psi(\brr).
\EEA
With this notation, the identification of the polarization and the spatial DoFs, represented by the Jones vector $\left[ A_0 , A_1 \right]^T$ and the scalar field $\psi(\brr)$, respectively, becomes straightforward.

For a field of the form (\ref{PW_field_Jones}), polarization and spatial DoFs are said to be \emph{separable} \cite{Luis} because the expression of $\bE(\brr)$ is factorizable in the product of a single \emph{space independent} vector, and a single scalar field. This mathematical property  reflects the  absence of physical coupling between polarization and spatial DoFs. However, in general, polarization and spatial DoFs can be coupled and when this happens the  factorizable  representation (\ref{PW_field_Jones}) is no longer valid. Consider, for example, the electric field of a beam of light \emph{nonuniformly} polarized in the $(x,y)$ plane, which can be expressed as
\BEA
\label{eq:gen_field}
\bE(\brr)=A_{00}\, \bhe_x \psi_{10}(\brr)+A_{01}\, \bhe_x \psi_{01}(\brr)+A_{10} \, \bhe_y \psi_{10}(\brr)+A_{11}  \, \bhe_y \psi_{01}(\brr),
\EEA
where $\psi_{mn}(\brr)$, with $m,n \in \{0,1,2,\ldots\}$, is the Hermite-Gauss (HG) solution of the paraxial wave equation of order $N=m+n$ \cite{Siegman} and $A_{ij}$ denotes a complex amplitude of the field, with $i,j \in \{0,1\}$.
In the Jones notation, (\ref{eq:gen_field}) takes either of the two following forms:
\numparts
\BEA
\bE(\brr)&= &  \left[\begin{array}{c}
 \!   \! A_{00}\psi_{10}(\brr) +  A_{01}\psi_{01}(\brr)\!  \! \\
  \! \!  A_{10}\psi_{10}(\brr) + A_{11}\psi_{01}(\brr)  \!  \!
\end{array}\right] \label{eq:gen_field_Ja} \\
&= &  \left[\begin{array}{c}
 \!   \! A_{00} \!  \! \\
  \! \!  A_{10}  \!  \!
\end{array}\right]\!  \psi_{10}(\brr) + \left[\begin{array}{c}
 \!   \! A_{01}\!  \! \\
  \! \!  A_{11}\!  \!
\end{array}\right] \! \psi_{01}(\brr). \label{eq:gen_field_Jb}
\EEA
\endnumparts

By writing the electric field as in \eref{eq:gen_field_Ja}, it follows that the beam has a nonuniform polarization pattern because the Jones vector varies with the position vector $\brr$. On the other hand, when $\bE(\brr)$ is written in the form \eref{eq:gen_field_Jb}, it appears evident that polarization and spatial DoFs are now \emph{nonseparable}, or entangled, because one needs two coordinate-independent Jones vectors and two independent scalar fields, $\psi_{10}$ and $\psi_{01}$,\footnote{The HG-modes $\psi_{mn}(\brr)$ are independent in the sense that they are orthonormal with respect to the spatial scalar product defined as $\int \! \!\int  \psi_{mn}^*(\brr) \psi_{m'n'}(\brr) \, \mathrm{d}x\,\mathrm{d}y = \delta_{m m'} \delta_{n n'}$, with $m,m',n,n' \in \{0,1,2,\ldots\}$.} to represent the electric field.

%
%
\begin{figure}[!ht]
\begin{center}
 \includegraphics[width=15.2truecm]{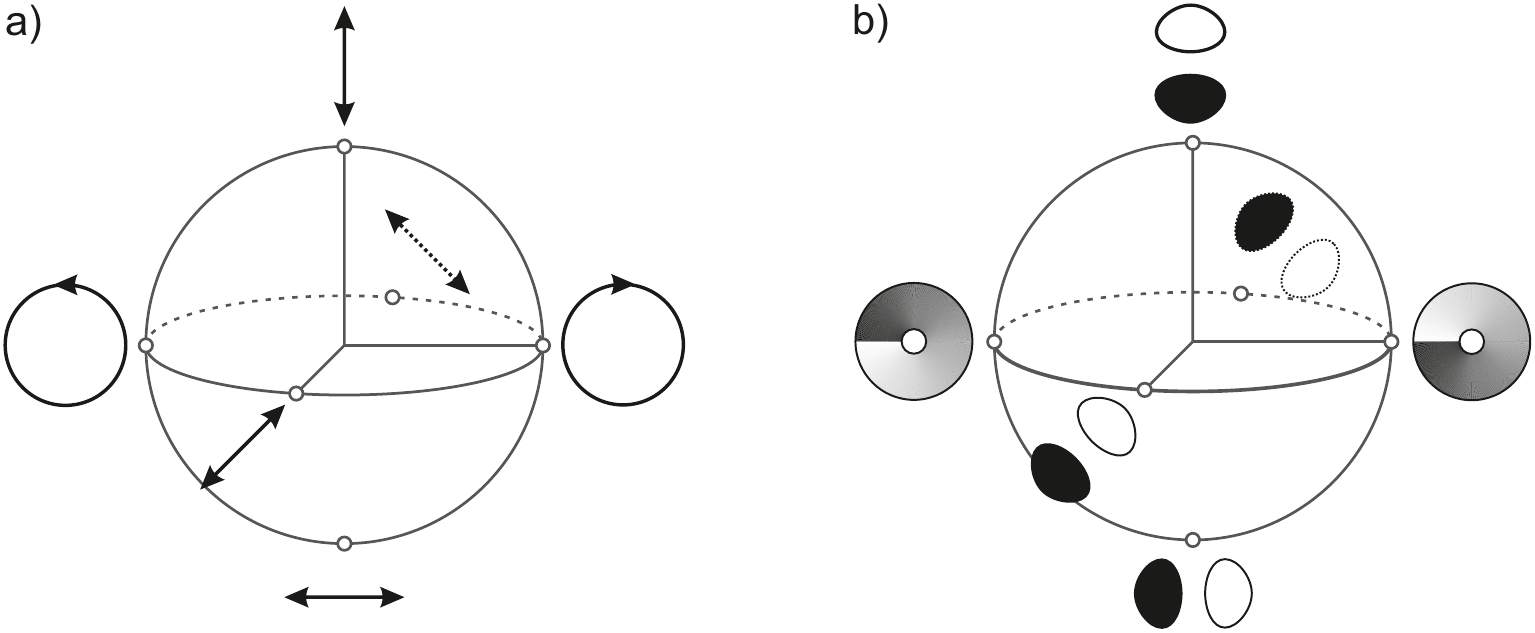}
\caption{\label{fig.poincare_sphere} Schematic visualization of a) the polarization Poincar\'e sphere representation of the binary Hilbert space $\mathscr{H}_\mathrm{pol}$ \cite{Damask} and b) Poincar\'e sphere representation of first-order spatial modes Hilbert space $\mathscr{H}_\mathrm{spa}$ \cite{Padgett}.}
\end{center}
\end{figure}
\begin{figure}[!ht]
\begin{center}
 \includegraphics[width=15.7truecm]{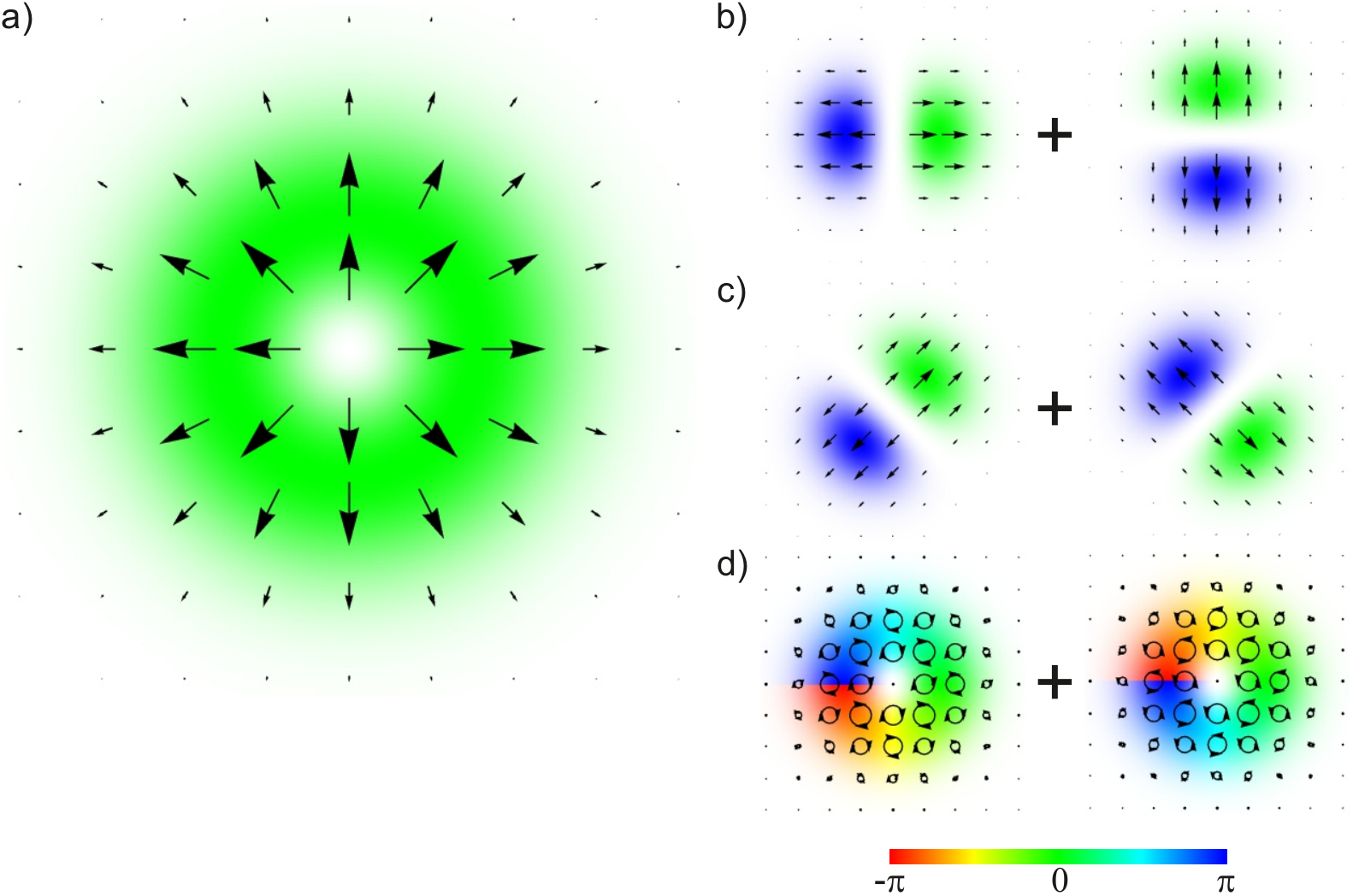}
\caption{\label{fig.mode_decompostion} a) Visual representation of the intensity distribution and polarization pattern of the radially polarized beam of light \eref{eq:psi_b}. b - d) Possible equivalent decompositions of the beam 
\eref{eq:psi_b}: The color scale (bottom) gives the phase of the electric field. b) Superposing the electric fields of a HG mode $\psi_{10}$  with horizontal polarization $\be_x$ and a HG mode $\psi_{01}$ with vertical polarization $\be_y$ yields a radially polarized pattern: $\bE = (\be_x \psi_{10} + \be_y \psi_{01})/\sqrt{2}$. c) Superposition of diagonal HG modes $\psi_\pm=(\psi_{10}\pm\psi_{01})/\sqrt{2}$ with diagonal polarizations $\be_\pm=(\be_x\pm\be_y)/\sqrt{2}$ also produces a radially polarized beam: $\bE = (\be_{+} \psi_{+} + \be_{-} \psi_{-})/\sqrt{2}$. d) Decomposition of a radially polarized beam into circular spatial modes $\psi_L=(\psi_{10}+\rmi\,\psi_{01})/\sqrt{2}$, $\psi_R=(\psi_{10}-\rmi\,\psi_{01})/\sqrt{2}$ and circular polarizations $\be_L=(\be_x+\rmi\,\be_y)/\sqrt{2}$, $\be_R=(\be_x-\rmi\,\be_y)/\sqrt{2}$ components: $\bE = (\be_L \psi_{R} + \be_R \psi_{L})/\sqrt{2}$.}
\end{center}
\end{figure}

%
%
Mathematically speaking, occurrence of entanglement requires an expression to be written as the sum of tensor products of two or more vectors belonging to different vector spaces. This is precisely what occurs in \eref{eq:gen_field_Jb}, where we have a polarization vector space and a spatial vector space, which can be represented by the polarization Poincar\'e sphere \cite{Damask} and the first-order spatial modes Poincar\'e sphere \cite{Padgett}, respectively, shown in figure \ref{fig.poincare_sphere}. 
This qualitative discussion may be made more quantitative by considering a radially polarized beam of light
 as a specific example that can be represented by (\ref{eq:gen_field}),
\BEA
\label{eq:psi_b}
\bE(\brr)=\frac{1}{\sqrt{2}}\Big[\bhe_x \, \psi_{10}(\brr) + \bhe_y \,\psi_{01}(\brr)\Big],
\EEA
whose characteristics are illustrated in figure \ref{fig.mode_decompostion}.
%
%
%
%
One can obtain \eref{eq:psi_b} from \eref{eq:gen_field} by putting $A_{00}=1/\sqrt{2}=A_{11}$ and $A_{01}=0=A_{10}$.
This suggests, as shown with great detail in \cite{Luis,Holleczek}, that
it is possible to represent a radially polarized beam by an abstract $4$-dimensional vector, henceforth denoted with the ket $|E \rangle  \doteq \left[ A_{00} , A_{01} , A_{10} , A_{11}\right]^T$, living in a $4$-dimensional  two-qubit Hilbert space: $|E \rangle \in\mathscr{H}_2=\mathscr{H}_\mathrm{pol} \otimes \mathscr{H}_\mathrm{spa}$. Here $\mathscr{H}_\mathrm{pol}=\mathrm{span}\{\bhe_x,\bhe_y\}$ denotes the polarization-qubit space and $\mathscr{H}_\mathrm{spa}=\mathrm{span}\{\psi_{10}(\brr),\psi_{01}(\brr)\}$ indicates the spatial-qubit space. We identify the standard basis for the polarization qubit with horizontal and vertical polarization states, and the standard basis for the spatial qubit with HG modes
of order $n+m = 1$, namely
\numparts
\BEA
\ket{0}_\mathrm{pol}\ce\bhe_x, & \qquad  \ket{1}_\mathrm{pol}\ce\bhe_y, \label{eq:basisP} \\
\ket{0}_\mathrm{spa}\ce\psi_{10}(\brr), & \qquad  \ket{1}_\mathrm{spa}\ce\psi_{01}(\brr). \label{eq:basisS}
\EEA
\endnumparts
We can  write a complete orthonormal basis of $\mathscr{H}_2$ in the form of a tensor product: $|i,j \rangle  = |i \rangle_\mathrm{pol} \otimes |j \rangle_\mathrm{spa}$ with $i,j \in \{ 0,1\}$, where the first index $i$ marks the polarization qubit  and the second index $j$ the spatial one, namely
\BEA
\label{eq:basis2}
\ket{0,0}\ce \bhe_x \psi_{10}(\brr), \qquad   \qquad \ket{0,1}\ce \bhe_x \psi_{01}(\brr), \nonumber \\
\ket{1,0}\ce \bhe_y \psi_{10}(\brr), \qquad   \qquad \ket{1,1}\ce \bhe_y \psi_{01}(\brr).
\EEA
 With this notation we can represent the electric field \eref{eq:gen_field} as a vector in $\mathscr{H}_2$:
\BEA
\label{eq:gen_field_ket}
|E \rangle = A_{00}\, \ket{0,0}+A_{01}\, \ket{0,1}+A_{10} \, \ket{1,0}+A_{11}  \, \ket{1,1},
\EEA
and a radially polarized beam can be described by the ket
\BEA
\label{eq:psi_c}
\ket{E}=\frac{1}{\sqrt{2}}\left(\ket{0,0}+\ket{1,1}\right) .
\EEA
This representation for the radially polarized beam is formally equivalent (isomorphic) to a Bell state of two qubits \cite{NielsenBook}. Hence, the polarization and spatial DoF may be treated as two qubits that are classically entangled.  However, the key difference between \eref{eq:psi_c} and a bona fide quantum optical Bell state is that in the latter the two qubits are encoded in the polarization DoFs of \emph{two} separated photons (inter-system entanglement), while in  \eref{eq:psi_c} the two qubits are encoded in the polarization and the spatial binary DoFs of a \emph{single}  light beam (intra-system entanglement).
%
%
\section{Stokes parameters and the Liouville representation of a quantum state}
The representations \eref{eq:gen_field} and \eref{eq:gen_field_ket} for the electric field of a light beam are adequate as long as one is concerned with
detection schemes that can resolve \emph{both} polarization and spatial DoFs.
When this is not the case, as in the conventional Mueller matrix polarimetry, it becomes necessary to introduce a more general representation, namely the $4 \times 4$ coherency matrix $\rho$ of the field, defined in terms of the electric field amplitudes $A_{ij}$ as
\BEA\label{eq:gen_cov}
|E \rangle\langle E| \ce \rho =
 \left[\begin{array}{cccc}
 A_{00}A_{00}^* & A_{00}A_{01}^* & A_{00}A_{10}^* & A_{00}A_{11}^* \\
 A_{01}A_{00}^* & A_{01}A_{01}^* & A_{01}A_{10}^* & A_{01}A_{11}^* \\
 A_{10}A_{00}^* & A_{10}A_{01}^* & A_{10}A_{10}^* & A_{10}A_{11}^* \\
 A_{11}A_{00}^* & A_{11}A_{01}^* & A_{11}A_{10}^* & A_{11}A_{11}^* \\
\end{array}\right].
\EEA
Of course, at this stage \eref{eq:gen_cov} still contains the same amount of information as \eref{eq:gen_field_ket}.
As a specific example, the coherency matrix of
the radially polarized beam \eref{eq:psi_c} can be simply written as
\BEA
\label{eq:gen_cov_radpol}
\rho =\frac{1}{2}
\left[\begin{array}{cccc}
 1 & 0 & 0 & 1 \\
 0 & 0 & 0 & 0 \\
 0 & 0 & 0 & 0 \\
 1 & 0 & 0 & 1
\end{array}\right].
\EEA

Suppose now to have a detection scheme that is not capable to resolve the spatial DoFs. In this case, \eref{eq:gen_field_ket} would furnish redundant information about the spatial DoFs that is not at our disposal. However, a proper representation of the beam can be then obtained from \eref{eq:gen_cov} by tracing out the unobservable spatial DoFs. In this manner one obtains the  reduced  $2 \times 2$ polarization coherency matrix $\rho_\mathrm{\,pol} $ that encodes all the available information about the polarization of the light beam, irrespective of the spatial DoFs:
\BEA
\label{eq:rho_pol}
\rho \rightarrow \rho_\mathrm{\,pol} & = & \tr_\mathrm{spa}\left( \rho \right) \nonumber \\
& = & \sum_{i=0}^1 {}_\mathrm{spa}\langle i | \rho | i \rangle_\mathrm{spa} \nonumber \\
& =  & A A^\dagger,
\EEA
where $\tr_\mathrm{spa}\left( \ldots \right)$ denotes the trace with respect to the spatial DoFs and the spatial kets $\ket{0}_\mathrm{spa}$ and $\ket{1}_\mathrm{spa}$ are defined in \eref{eq:basisS}. Here
\BEA
A = \left[\begin{array}{cc}
 A_{00} & A_{01} \\
 A_{10} & A_{11} 
\end{array}\right],
\EEA
is a $2 \times 2$ matrix whose elements are the coefficients $A_{ij}$ of the ket expansion \eref{eq:gen_field_ket}.
In a similar manner, one can calculate the  reduced  $2 \times 2$ spatial coherency matrix $\rho_\mathrm{\,spa} $ that encodes all the available information about the spatial modes of the light beam, irrespective of the polarization:
\BEA
\label{eq:rho_spa}
\rho_\mathrm{\,spa} & = & \tr_\mathrm{pol}\left( \rho \right) \nonumber \\
& =  & \left(A^\dagger A \right)^T ,
\EEA
where $\tr_\mathrm{pol}\left( \ldots \right)$ denotes the trace with respect to the polarization DoFs.

From the definition \eref{eq:rho_pol} it follows that $ \rho_\mathrm{\,pol} $ is Hermitian and positive semidefinite. Therefore, it admits a Liouville representation of the form
\BEA
\label{eq:rho_pol_stokes}
\rho_\mathrm{\,pol}  = \frac{1}{2} \sum_{\mu =0}^3 S_\mu \sigma_\mu,
\EEA
where the coefficients $S_\mu$ are real numbers and the set $\{ \sigma_\mu\}_{\mu=0}^3$ of $2 \times 2$ Hermitean matrices forms a complete basis of observables \cite{Balian}. The factor $1/2$ in front of \eref{eq:rho_pol_stokes} is conventional. In classical polarization optics, the coefficients $S_\mu$ are known as the Stokes parameters of the beam \cite{MandelBook} and the basis  set $\{ \sigma_\mu\}_{\mu=0}^3$ is constituted  by the four Pauli matrices
\BEA
\label{eq:pauli_matrices}
&\sigma_0=
\left[\begin{array}{cc}
 1 & 0\\
 0 & 1
\end{array}\right],\qquad\qquad
&\sigma_1=
\left[\begin{array}{cc}
 0 & 1\\
 1 & 0
\end{array}\right],\nonumber\\
&\sigma_2=
\left[\begin{array}{cc}
 0 & -\rmi\\
 \rmi & 0
\end{array}\right],\qquad\qquad\
&\sigma_3=
\left[\begin{array}{cc}
 1 & 0\\
 0 & -1
\end{array}\right],
\EEA
which are orthogonal with respect to the scalar product defined as $\tr \left( \sigma_\mu \sigma_\nu \right) = 2 \delta_{\mu \nu}$. From this property and the definition \eref{eq:rho_pol_stokes} it follows that 
\BEA\label{Smu}
S_\mu = \tr \left( \rho_{\,\mathrm{pol}} \sigma_\mu \right).
\EEA

For the radially polarized beam (\ref{eq:psi_c}), $A = \mathcal{I}_2/\sqrt{2}$ and $\rho_{\,\mathrm{pol}} = \mathcal{I}_2/2 =  \rho_\mathrm{\,spa}$, where $\mathcal{I}_2$ denotes the $2 \times 2$ identity matrix.
In the language of classical polarization optics, this means that the radially polarized beam is completely unpolarized: $\bS=[S_0,S_1,S_2,S_3]^T = [1,0,0,0]^T$.
{
This observation may appear confusing because \eref{eq:gen_field_Ja} and figure \ref{fig.mode_decompostion} show that a radially polarized beam possesses a well defined \emph{local}, i.e. defined in each point $(x,y)$ of its transverse spatial profile, Jones vector given by}
\BEA\label{eq:gen_field_Ja_RP}
\bE(\brr) =   \frac{1}{\sqrt{2}}\left[\begin{array}{c}
 \!   \! \psi_{10}(\brr) \!  \! \\
  \! \!  \psi_{01}(\brr)  \!  \!
\end{array}\right].
\EEA
{However,
$\rho_{\,\mathrm{pol}}$ is} obtained from $\rho$ after tracing out the spatial DoFs. From a physical point of view, this corresponds to measuring the \emph{global} Stokes parameters of the beam, as a whole, with ``bucket'' detectors that integrate the intensity of light over all the cross section of the beam.
A similar situation is encountered for photon-pairs in a Bell state: Although the polarization of the two-photon state is perfectly defined (pure state), each of the two photons, when observed separately, appears as completely unpolarized (mixed state) \cite{Lehner}.
\section{Mueller matrix polarimetry}
Typically, in a conventional polarimetry setup, an either transmissive or scattering material sample (the object) is illuminated with a light beam (input beam) that, as a result of the interaction with the object, emerges transformed (output beam).
In this section we will study how radially polarized beams transform under the action of a polarization-affecting optical element, having in mind the final goal of measuring the Mueller matrix of the latter. From a mathematical point of view, here we consider {local} linear transformations of the form $T_\mathrm{pol} \otimes T_\mathrm{spa}$, namely transformations that act on each DoF separately, where $T_d: \mathscr{H}_d \rightarrow \mathscr{H}_d$ with $d\in \{\mathrm{pol},\mathrm{spa}\}$ is a $2 \times 2$ complex matrix known as Jones matrix in polarization optics. As in this work we are concerned with optical elements affecting polarization DoFs solely, henceforth we assume $T_\mathrm{spa} = \mathcal{I}_2$, and we will omit the subscript ``pol'' in $T_\mathrm{pol}$.

Under the action of $T$, the generic ket \eref{eq:gen_field_ket} transforms as
\BEA
\label{eq:ket_tran_1}
\ket{E} \rightarrow \ket{E'} & = & \left( T \otimes \mathcal{I}_2 \right) \ket{E}  \nonumber \\
 & = & \sum_{i,j=0 }^{1} A'_{ij} \ket{i,j},
\EEA
with $A' \ce T A$.
 The transformation \eref{eq:ket_tran_1} links the amplitudes $A_{ij}$ of the input beam to the amplitudes $A_{ij}'$ of the output beam. However, in real-world experiments intensities, rather than amplitudes, are measured.
Therefore, it becomes necessary to specify the type of intensity measurements actually
performed upon the output beam. According to whether the detectors are or are not insensitive to the spatial DoFs of the beam, one deals with either
(a) single-DoF polarimetry or (b) two-DoF polarimetry. Case (a) coincides with the conventional Mueller matrix polarimetry,  while case (b) gives the novel detection scheme that we propose here. Let us shortly review case (a) first.
\subsection{Single-DoF polarimetry}
From \eref{eq:ket_tran_1} and the definition \eref{eq:rho_pol} it follows that $\rho_{\, \mathrm{pol}}$ transforms under $T$ as
\BEA
\label{eq:ket_tran_1b}
\rho_{\, \mathrm{pol}} \rightarrow \rho{\,'}_{\! \mathrm{pol}} = A' {A'}^\dagger= T A A^\dagger T^\dagger.
\EEA
Suppose we prepare sequentially the input beam in four different polarization states labeled by the index $\alpha\in\{0,1,2,3\}$. For example, $\alpha=0$ may denote horizontal polarization, $\alpha=1$ vertical polarization, $\alpha=2$ diagonal polarization and  $\alpha=3$ left-circular polarization.
 Then, in the Liouville representation \eref{eq:rho_pol_stokes} and by using the definition \eref{Smu}, the  transformation \eref{eq:ket_tran_1b} can be written, for each different input beam labeled by the index $\alpha$,  as
\BEA\label{eq:transf_stokes_2}
S'_\mu(\alpha) = \sum_{\nu=0}^3 M_{\mu \nu} \, S_\nu(\alpha), \qquad \mu,\alpha\in\{0,1,2,3\},
\EEA
where $S_\nu(\alpha)$ and $S'_\mu(\alpha)$ denote the Stokes parameters of the input and output beams, respectively, and the $16$ real numbers
\BEA\label{eq:MM}
M_{\mu \nu} = \frac{1}{2}\tr \left(\sigma_\mu T \sigma_\nu T^\dagger \right),
\EEA
 are the (unknown) elements of the sought Mueller matrix $M$.\footnote{ {When the object is a  depolarizing optical element, then \eref{eq:MM} must be replaced with $M_{\mu \nu} = \tr (\overline{\sigma_\mu T \sigma_\nu T^\dagger })/2$, where the $\overline{\mathrm{overline}}$ symbol denotes average over a stochastic set \cite{Damask}. For the sake of clarity, in the remainder we will consider only non-depolarizing optical elements.}}
Then, \eref{eq:transf_stokes_2} may be seen as a linear system of $16$ equations and $16$ unknowns that can be easily solved, for example,  by defining the two $4 \times 4$ matrices $V$ and $V'$ as: $[V]_{\nu \alpha} \ce S_\nu(\alpha)$ and $[V']_{\mu \alpha} \ce S'_\mu(\alpha)$. This permits to rewrite \eref{eq:transf_stokes_2} in the simple matrix form  $V' = M V$,  and the Mueller matrix $M$ can be finally evaluated as
\BEA\label{eq:transf_stokes_3}
 M = V' V^{-1},
\EEA
providing that $\det \left( V \right)\neq 0$.

This is
the essence of conventional Mueller matrix polarimetry \cite{Bickel}. Of course, in a situation where experimental errors may occur, the simple linear inversion algorithm \eref{eq:transf_stokes_3} often does not suffice and more sophisticated inversion methods must be used instead \cite{Jeune,Aiello2006}.
However, the lesson to be learned here is that conventional Mueller matrix polarimetry needs the input beam to be sequentially prepared in, at least, four different polarization states to gain the complete information about the object. Conversely, we are going to show soon how the same amount of information can be obtained by probing the object only once with a radially polarized beam.
%
%
\subsection{Two-DoF polarimetry}
We now consider a detection scheme that is capable to resolve \emph{both} the polarization and  the spatial DoFs. The complete coherency matrix
\eref{eq:gen_cov} can also be written in a Liouville form similar to \eref{eq:rho_pol_stokes}  as
\BEA
\label{eq:rho_pol_stokes_2}
\rho  = \frac{1}{4} \sum_{\mu,\nu =0}^3 S_{\mu \nu} \left( \sigma_\mu \otimes \sigma_\nu \right),
\EEA
where we have defined the two-DoFs Stokes parameters as \footnote{Here we are using the standard properties of the direct product of matrices: $\left( A \otimes B\right) \left(C \otimes D \right) = A C \otimes B D$ and  $\tr \left( A \otimes B\right) =\tr \left( A\right)\tr \left( B\right)$. }
\BEA
\label{eq:stokes_correlation}
S_{\mu\nu}=\tr \left[ \rho  \left(\sigma_\mu\otimes\sigma_\nu \right)\right].
\EEA
These quantities are the classical optics analogue of the two-photon Stokes parameters introduced in \cite{James,AbouraddyOC}. However, while in \cite{AbouraddyOC} the two polarization qubits are encoded in two \emph{separated} photons, in our case the polarization qubit and the spatial qubit are encoded in the \emph{same} radially polarized beam of light.
Therefore, the two-DoFs Stokes parameters give the \emph{intrabeam} correlations between polarization and spatial DoFs \cite{Loudon}. In order to measure these correlations, one needs a detection scheme capable to resolve both DoFs. Such an experimental apparatus will be studied in the next section.
For the radially polarized beam represented by
\eref{eq:gen_cov_radpol}, the two-DoFs Stokes parameters take the particularly simple form 
\BEA
S_{\mu \nu} = \lambda_\mu \delta_{\mu \nu},\qquad\mathrm{where}\qquad\{\lambda_\mu\}_{\mu=0}^3 = \{1,1,-1,1\}.
\EEA
From \eref{eq:ket_tran_1} it follows that, under the action of $T$, \eref{eq:gen_cov_radpol} transforms as
\BEA\label{eq:stokes_mueller_0}
\rho \rightarrow \rho{\,'}&= \; (T\otimes \mathcal{I}_2)\rho \,(T^\dagger\otimes \mathcal{I}_2)\nonumber \\
&= \; \frac{1}{4}\sum_{\mu=0}^3 \lambda_\mu (T\sigma_\mu T^\dagger)\otimes\sigma_\mu.
\EEA
Substituting \eref{eq:stokes_mueller_0} into \eref{eq:stokes_correlation} yields
\BEA
\label{eq:stokes_mueller}
S'_{\mu\nu}&= \tr\left[\rho{\,'}\!\left(\sigma_\mu\otimes\sigma_\nu \right)\right]\nonumber\\
&= \frac{1}{4}\sum_{\alpha=0}^3 \lambda_\alpha \tr\bigl( \sigma_\mu T\sigma_\alpha T^\dagger\bigr)
\tr\bigl(\sigma_\alpha\sigma_\nu\bigr)
\nonumber \\
&=  M_{\mu\nu} \lambda_\nu,
\EEA
where \eref{eq:MM} has been used in the last line. Since either $\lambda_\nu=1$ for $\nu\in\{0,1,3\}$ or $\lambda_\nu=-1$ for $\nu=2$, from \eref{eq:stokes_mueller} it follows that the two-DoF Stokes parameters furnish a direct measure of the Mueller matrix elements:
\BEA
\label{eq:result}
M_{\mu\nu}=
\cases{
-S'_{\mu\nu}, & for $\nu=2$,\\
\phantom{-}S'_{\mu\nu}, & for $\nu\neq2$.\\
}
\EEA
This shows that the Mueller matrix of an object can be obtained from the measurement of the $16$ two-DoFs Stokes parameters $S'_{\mu\nu}$, with a single radially polarized input beam, allowing to perform single-shot full polarimetry.
We remark that the above derivation relies upon the assumption that the optical properties of the object do not vary over the cross section of the input beam, namely that $T$ is independent of $x$ and $y$.

This interesting result can be understood as an effect of postselection on an entangled state. In a single-DoF polarimetry setup, the polarization state of the input beam is \emph{preselected} before the interaction with the object, as shown in \eref{eq:transf_stokes_2}. Consequently, the object can be probed by only a single polarization state at a time. Vice versa, in our  two-DoF polarimetry scheme the polarization state of the input beam is \emph{postselected} after the interaction with the object via the two-DoF correlations measurements. Therefore, the object is probed, at once, by all possible polarization states carried by the radially polarized beam. This magic is made possible by the entangled structure \eref{eq:psi_b} or \eref{eq:psi_c} of the beam: projecting the output beam on a specific spatial mode uniquely determines, a posteriori, the polarization of the input beam, which may be either linear, diagonal or circular as shown in figure \ref{fig.mode_decompostion}.
%
%
\section{Real-time single-shot Mueller matrix polarimetry}
\begin{figure}[!ht]
\begin{center}
 \includegraphics[width=15.6truecm]{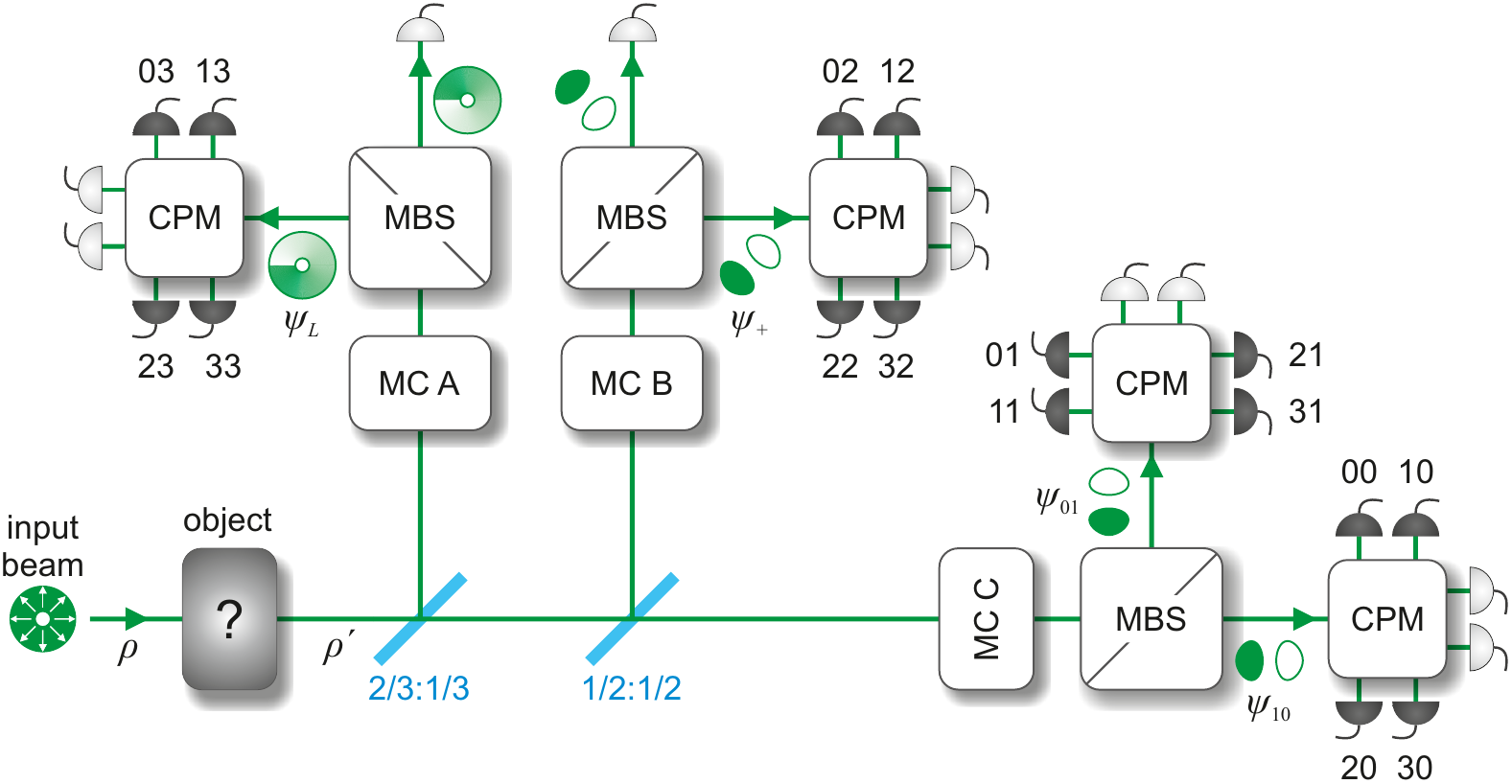}
\caption{\label{fig.full_setup} Schematic setup for real-time single-shot Mueller matrix polarimetry. The radially polarized input beam $\rho$ propagates through a material sample (object) whose Mueller matrix must be determined. The light $\rho'$ transmitted by the sample is split by polarization maintaining beam splitters (BSs) in three identical beams and sent through mode converters (MCs) followed by mode beam splitters (MBSs). The ratios of transmission ($t$) and reflection ($r$) coefficients of the BSs $|t|^2:|r|^2$ are indicated in the figure. Each combination of a MC with a MBS effectively projects the entering beam on the specific spatial mode represented schematically in the figure. These operations postselect the polarization of the input beam as explained in the text. The four output ports of the three MBSs projecting onto the spatial modes $\{\psi_{10},\psi_{01},\psi_+,\psi_L\}$ (also denoted with $\{\psi_0,\psi_1,\psi_2,\psi_3 \}$),
are coupled to four distinct conventional polarization measurement setups (CPMs). Each CPM delivers the intensities of the four polarization components $\{\bhe_x,\bhe_y,\bhe_+,\bhe_L\}$ (also denoted with $\{\be_0,\be_1,\be_2,\be_3 \}$), per each of the four entering beams $\psi_{10}$, $\psi_{01}$, $\psi_+$ or $\psi_L$. The detectors labeled with the polarization-spatial indexes $\alpha\beta$ with $\alpha, \beta \in \{ 0,1,2,3\}$ in the figure, return the $4\times4=16$ intensities $I_{\alpha\beta}$, where, for example, $I_{10}=\bra{\be_y,\psi_{10}}\rho'\ket{\be_y,\psi_{10}}$. From these intensities the two-DoF Stokes parameters $S_{\mu\nu}$ and the Mueller matrix can be completely determined via \eref{V.110} and \eref{eq:result}. 
In a setup that suffers from experimental imperfections, the intensities obtained by the 10 additional detectors displayed in light gray may be needed.
Physical implementations of the MCs, MBS and CPM are shown in figures \ref{fig.MCP_MBS} and \ref{fig.CPM}.}
\end{center}
\end{figure}

\begin{figure}[!ht]
\begin{center}
 \includegraphics[width=13.8truecm]{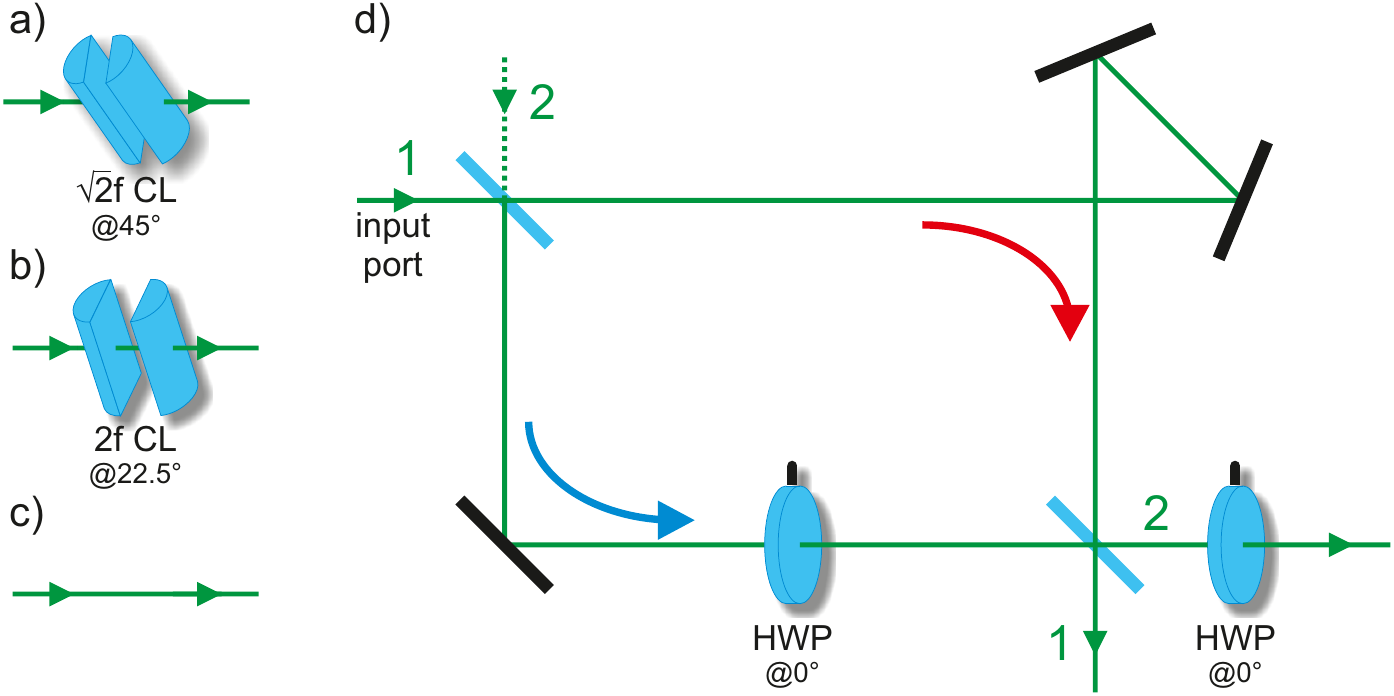}
\caption{\label{fig.MCP_MBS} Physical implementation of mode converters (MCs) and the mode beam splitter (MBS) used in figure \ref{fig.full_setup}. a) \textsf{MC A} comprises a pair of identical cylindrical lenses at a distance of $\sqrt{2}f$ rotated by $45^\circ$ with respect to the $x$-axis. b) \textsf{MC B} consists of a pair of identical cylindrical lenses separated by twice their focal length $f$ and oriented at an angle of $22.5^\circ$ with respect to the $x$-axis. a) \textsf{MC C} contains no optical elements. d) The MBS is realized with a Mach-Zehnder interferometer with an additional mirror in one arm and an additional half-wave plate (HWP) in the other arm with its fast axis oriented along the direction of horizontal polarization. The output port 2 of the interferometer is coupled to another HWP oriented identically to the first one. The presence of the second HWP becomes important when the MBS is coupled to a CPM.}
\end{center}
\end{figure}

\begin{figure}[!ht]
\begin{center}
 \includegraphics[width=12.8truecm]{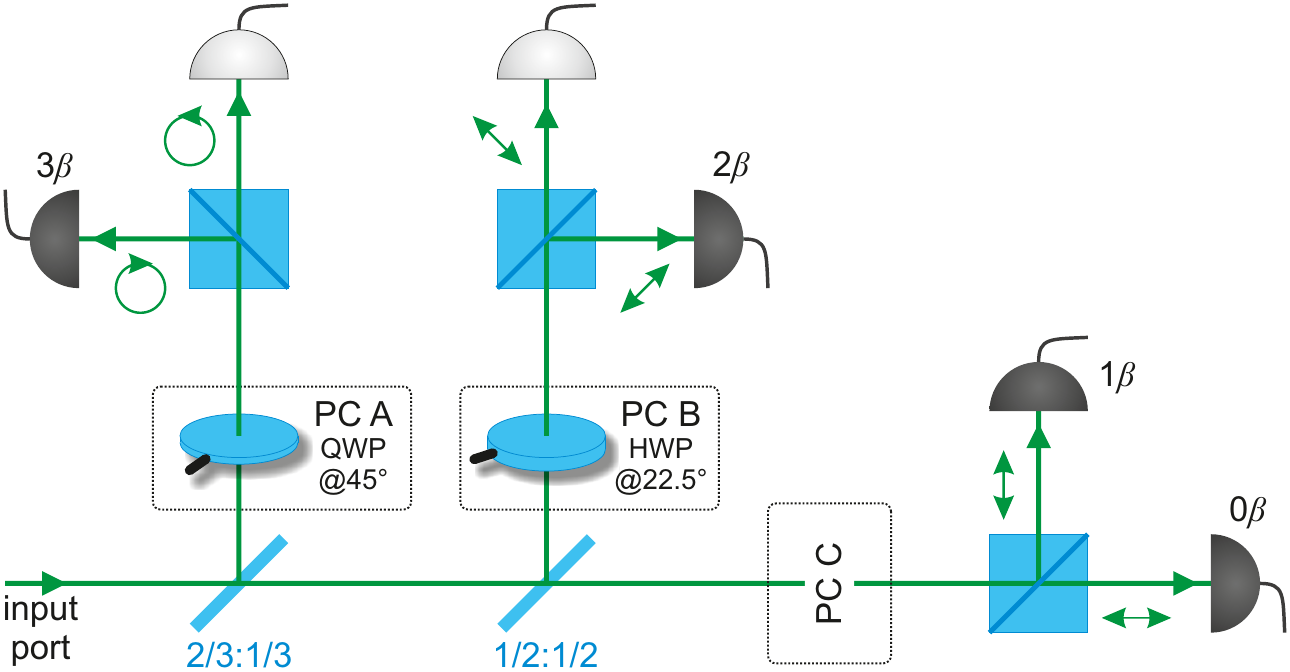}
\caption{\label{fig.CPM} Physical implementation of the conventional polarization measurement setups (CPMs) used in figure \ref{fig.full_setup}. The input beam is split into three identical beams by two polarization maintaining beam splitters. Each of the three beams passes through a polarization converter (PC) followed by an polarizing beam splitter (PBS). \textsf{PC A} and \textsf{PC B} consist of a quarter-wave plate (QWP) and a half-wave plate (HWP), respectively. \textsf{PC C} is made from empty space. Complete information on the polarization state of the beam entering the input port can, in principle, be inferred from the measurements delivered by the 4 detectors labeled by the polarization-spatial indexes $\{ 0 \beta, 1 \beta, 2 \beta, 3 \beta\}$, where $\beta\in\{0,\dots,3\}$ is the spatial mode index. However, in an experimental realization with unavoidable losses, the intensities measured by the 2 additional detectors displayed in light gray may be needed.}
\end{center}
\end{figure}

In this section we propose a feasible experimental scheme  for real-time single-shot  Mueller matrix polarimetry. The measurement setup is illustrated in figures \ref{fig.full_setup}-\ref{fig.CPM}. The procedure we present here is an extension of the conventional polarization measurement  technique to beams of light with coupled polarization and spatial DoFs.
According to  \eref{eq:result}, the fundamental quantities to estimate are the sixteen two-DoF Stokes parameters $S_{\mu\nu}$ that contain all the information about the Mueller matrix. 
%
%
%
%

The procedure is as follows: a radially polarized  beam of light (the probe) is sent through
a material sample (the object) whose Mueller matrix has to be  determined.  
Then, the idea is to first project the beam transmitted across the sample onto the four independent spatial modes $\{\psi_{10},\psi_{01},\psi_+,\psi_L\}$ (also denoted with $\{\psi_0,\psi_1,\psi_2,\psi_3 \}$.)%
\footnote{Diagonal/antidiagonal spatial modes are defined as $\psi_{\pm} = \left( \psi_{10} \pm  \psi_{01}\right)/\sqrt{2}$ and  left/right-circular spatial modes are written as $\psi_{L}= \left( \psi_{10} + \rmi \, \psi_{01}\right)/\sqrt{2}$ and $\psi_{R}= \left( \psi_{10} - \rmi \, \psi_{01}\right)/\sqrt{2}$, respectively. Similarly, diagonal/antidiagonal polarization states are defined as $\be_\pm=(\be_x \pm \be_y)/\sqrt{2}$ and  left/right-circular polarization ones as $\be_L=(\be_x + \rmi\,\be_y)/\sqrt{2}$ and $\be_R=(\be_x - \rmi\,\be_y)/\sqrt{2}$, respectively.}
 These projections postselect the four independent polarization states $\{\bhe_x,\bhe_y,\bhe_+,\bhe_R\}$, as explained in detail in subsection \ref{sec51}. It is important to stress that we perform these operations without acting directly on the polarization DoFs, which are analyzed only in a subsequent stage. For this, the light transmitted by the sample is split in three identical beams of equal intensity, which then are sent to mode converters (MCs) and mode beam splitters (MBSs). The three MCs, denoted as \textsf{A}, \textsf{B} and \textsf{C} in figure \ref{fig.full_setup}, define in which basis  $(\psi_{L},\psi_{R})$, $(\psi_{+},\psi_{-})$ or $(\psi_{10},\psi_{01})$ the incoming beam is going to be measured. 
MC \textsf{A} that transforms the modes $(\psi_{L},\psi_{R})$ into $(\psi_{10},\psi_{01})$, is made of a $\pi/2$-converter, rotated by the angle $\theta=\pi/4$ with respect to the horizontal axis (see \ref{app:A}). MC \textsf{B}, which transforms the modes $(\psi_{+},\psi_{-})$ into $(\psi_{10},\psi_{01})$, is made of a $\pi$-converter, rotated by the angle $\theta=\pi/8$ with respect to the horizontal axis (see \ref{app:A}). MC \textsf{C} is made of empty space and does not change the modes. It can be shown \cite{Beijersbergen} that a $\pi$-MC can be physically realized with two identical cylindrical lenses separated by a distance $2f$ equal to two focal lengths ($2f$ CL), as shown in figure \ref{fig.MCP_MBS} a). Similarly, a $\pi/2$-MC is made from two identical cylindrical lenses separated by a distance $\sqrt{2}f$ ($\sqrt{2} f$ CL), as illustrated in figure \ref{fig.MCP_MBS}b). 
Each MC is coupled to a mode beam splitter (MBS), which splits up a beam into its $\psi_{10}$ and $\psi_{01}$ spatial components. The MBS is made of a modified Mach-Zehnder interferometer (MZ) with an extra mirror in one arm and a half-wave plate (HWP) in the other arm, followed by another HWP in one output port, as shown in figure \ref{fig.MCP_MBS} d) (see \ref{app:ms} and \cite{Sasada,Paz}). As a result of these transformations, the MBS placed behind MC \textsf{A} splits up the incoming beam into its circular spatial components, $\psi_L$  and $\psi_R$, MC \textsf{B} splits the beam into its diagonal and antidiagonal spatial components $\psi_+$ and $\psi_-$ and MC \textsf{C} into the $\psi_{10}$ and $\psi_{01}$ components. By selecting one of the outputs of MBS \textsf{A} and of MBS \textsf{B} and the two outputs of MBS \textsf{C}, one has access to the four spatial modes $\{\psi_{10},\psi_{01},\psi_+,\psi_L\}$. With this operation, we have physically acted only upon the spatial modes of the beam, and we will now analyze their polarization. This corresponds to a postselection of the polarization state of the probe beam. This is possible thanks to the entanglement between the polarization and the spatial DoFs.
%
%
%
%

The light exiting an output port of each MBS can be either directly detected, or
sent through a conventional polarization measurement (CPM) setup shown in figure \ref{fig.CPM}. 
A beam entering the CPM is split into three beams of equal intensity. Each beam passes through a polarization converter (PC) denoted with \textsf{A}, \textsf{B} and \textsf{C} in figure 3.  They are made from, respectively, \textsl{a}) a quarter-wave plate (QWP) whose fast axis is tilted by $\pi/4$ with respect to the horizontal direction; \textsl{b}) a HWP with the fast axis rotated by $\pi/8$ with respect to the horizontal direction; and \textsl{c}) empty space.
 Each PC is coupled to a polarizing beam splitter (PBS), which splits a beam into its  horizontal $\be_x$   and vertical  $\be_y$ polarization components. The three combinations  \textsf{A}, \textsf{B} and \textsf{C} of PCs and PBSs project the entering beams onto  three mutually unbiased pairs of polarization states, namely, \textsl{a}) horizontal/vertical: $\{\be_x, \, \be_y\}$; \textsl{b}) diagonal/antidiagonal:  $\{\be_+, \, \be_-\}$; and \textsl{c})  left/right-circular: $\{\be_L, \, \be_R\}$, respectively.
The intensity $I_{\alpha \beta}$ of the light projected in the state $
\be_\alpha\psi_\beta$ is recorded by a photo-detector that is identified  with the same pair of indexes $\alpha \beta: \,\alpha, \beta \in \{0,\dots,3\}$. Here the first index $\alpha$ marks the polarization qubit  and the second index $\beta$ the spatial one, as shown in  figure 1.
%
%
%
%
When four independent spatial modes are sorted $\{\psi_{10}, \, \psi_{01}, \, \psi_+, \, \psi_L\}\equiv \{\psi_0,\psi_1,\psi_2,\psi_3 \}$ by proper combinations of MCs and MBSs, and four independent polarization states $\{\be_x, \, \be_y, \, \be_+, \, \be_L\}\equiv \{\be_0,\be_1,\be_2,\be_3 \}$ are selected by convenient sequences of PCs and PBSs,  then the sixteen two-DoF Stokes parameters $S_{\mu \nu}$ can be entirely determined. In an ideal situation, the minimal number of detectors needed for this measurement is clearly $16$. However, in real-world experiments where uncontrollable losses may occur,  a maximal amount of $10$ additional detectors may be used to ensure proper normalization of all measured quantities. It should be noticed that also in conventional Mueller matrix polarimetry at least $16$ independent intensity measurements are required. The major advantage in our scheme is that we can perform the $16$ measurements 
at the same time, thus providing a ``real-time'', and potentially fast, Mueller matrix determination.

\subsection{Determining the two-DoF Stokes parameters}
\label{sec51}
In the remainder of this section, we will illustrate explicitly the two-DoF Stokes parameters measurement process putting particular emphasis into the postselection technique. For the sake of clarity, we will consider only the case of non-depolarizing (unknown) transmitting objects. More information about single- and two-qubit operations can be found in \ref{app:A}.

The radially polarized input beam \eref{eq:psi_c} can be represented by
\BEA\label{N10}
\ket{E} = \frac{1}{\sqrt{2}} \left(  \ket{\be_x}\ket{\psi_{10}} +  \ket{\be_y}\ket{\psi_{01}} \right),
\EEA
where the two-qubit basis \eref{eq:basis2}  has been  recast here in a more suggestive form. After interacting with the object,  characterized by the (unknown) Jones matrix $T$, the state of the input beam  is transformed according to:
\BEA\label{N20}
\ket{E} \stackrel{T}{\rightarrow} \ket{E'} = \frac{1}{\sqrt{2}} \Bigl[ \left( T \ket{\be_x} \right) \ket{\psi_{10}} + \left( T \ket{\be_y} \right)\ket{\psi_{01}} \Bigr].
\EEA
%
Using the decompositions of a radially polarized beam shown in figure \ref{fig.mode_decompostion},
the state  $\ket{E'}$ of the transmitted beam can be also written in the diagonal and circular mode bases as: 
\numparts
\BEA
\ket{E'} & = & \frac{1}{\sqrt{2}} \Bigl[ \left( T \ket{\be_+} \right) \ket{\psi_{+}} + \left( T \ket{\be_-} \right)\ket{\psi_{-}} \Bigr] \label{N30A} \\
 & = &  \frac{1}{\sqrt{2}} \Bigl[ \left( T \ket{\be_L} \right) \ket{\psi_{R}} + \left( T \ket{\be_R} \right)\ket{\psi_{L}} \Bigr]. \label{N30B}
\EEA
\endnumparts
The three combinations of MCs and MBSs  project the state $\ket{E'} $  onto the four independent modes $\psi_{10}, \, \psi_{01}, \, \psi_+, \, \psi_L$. These are  two-step operations: first a MC transforms the state $\ket{E'}$ into a chosen basis,  then a MBS projects the transformed state onto the ``linear'' basis $\{ \psi_{10}, \,\psi_{01}\}$. For example, from \eref{V.50} and \eref{N30A} it follows that the transformation performed by MC \textsf{B} (first step) produces
\BEA\label{N35}
\mathrm{MC}\;\mathsf{B} \; \longrightarrow \; U_\pi(\pi/8)\ket{E'} & = & \frac{- \rmi}{\sqrt{2}} \Bigl[ \left( T \ket{\be_+} \right) \ket{\psi_{10}} + \left( T \ket{\be_-} \right)\ket{\psi_{01}} \Bigr] .
\EEA
Then, in the second step the MBS projects the state onto $\psi_{01}$ and the result is:
\BEA\label{N37}
\mathrm{MBS} \; \longrightarrow \; \langle \psi_{01} |U_\pi(\pi/8)\ket{E'} & = & \frac{- \rmi}{\sqrt{2}} \,  T \ket{\be_-}.
\EEA
These projections provide postselection of the four input polarization states $\be_x, \, \be_y, \, \be_+, \, \be_R$, according to
\numparts
\BEA
\langle \psi_{10}| E'\rangle & = & \frac{1}{\sqrt{2}} \, T \ket{\be_x} , \label{N40A} \\
\langle \psi_{01}| E'\rangle & = & \frac{1}{\sqrt{2}} \, T \ket{\be_y} , \label{N40B} \\
\langle \psi_{+}| E'\rangle & = & \frac{1}{\sqrt{2}} \, T \ket{\be_-} , \label{N40C} \\
\langle \psi_{L}| E'\rangle & = & \frac{1}{\sqrt{2}} \, T \ket{\be_R} , \label{N40D}
\EEA
\endnumparts
where irrelevant overall phase factors have been omitted.
The  states (\ref{N40A}-\ref{N40D}) exiting four different ports of the three MBSs, are then analyzed by CPMs that allow to evaluate all the elements of the Jones matrix $T$. Consider, for example, \eref{N40A}. When this state is sent through a CPM, the following intensities can be measured:\footnote{As a technical remark, it should be noticed that the postselected set of polarization vectors $\{ \be_x, \, \be_y, \, \be_+, \, \be_R \} $ does not coincide with the analyser basis $\{ \be_x, \, \be_y, \, \be_+, \, \be_L \} $. However, this is not a problem as long as both set of vectors are linearly independent.}
\numparts
\BEA
I_{00} \equiv\left|\langle \be_x | \langle \psi_{10}| E'\rangle \right|^2 & = & \frac{1}{2} \, \left| \bra{\be_x} T \ket{\be_x} \right|^2 , \label{N50A} \\
I_{10} \equiv\left|\langle \be_y | \langle \psi_{10}| E'\rangle \right|^2 & = & \frac{1}{2} \, \left| \bra{\be_y} T \ket{\be_x} \right|^2 , \label{N50B} \\
I_{20} \equiv\left|\langle \be_+ | \langle \psi_{10}| E'\rangle \right|^2 & = & \frac{1}{2} \, \left| \bra{\be_+} T \ket{\be_x} \right|^2 , \label{N50C} \\
I_{30} \equiv\left|\langle \be_L | \langle \psi_{10}| E'\rangle \right|^2 & = & \frac{1}{2} \, \left| \bra{\be_L} T \ket{\be_x} \right|^2. \label{N50D}
\EEA
\endnumparts
When  the $16$ intensities $I_{\alpha \beta}$ are measured, eventually the $16$ two-DoF Stokes parameters $S_{\mu \nu}$ can be determined according to the formulas \eref{V.110} given in \ref{app:A2}.
\section{Conclusions}
%
%
%
In this work we have shown how to exploit {classical} entanglement in polarization metrology, by using radially polarized beams of classical light to perform {real-time} single-shot Mueller matrix measurements. Our main result is 
that the Mueller matrix elements are simply proportional to the two-DoF Stokes parameters that quantify the {intrabeam} correlations between polarization and spatial DoFs of a radially polarized beam.
The novelty of our approach is  that while the speed of conventional Mueller matrix measurements is limited by the need of probing the sample four {times} \emph{in sequence} with light of different polarization, in our setting the four probes are made \emph{in parallel} via a single radially polarized beam of light.
In conclusion, we have established a novel two-DoF polarimetry scheme, which is the classical wave analogue of two-photon polarimetry {\cite{AbouraddyOC}}. Our results generalise and extend to the classical optics regime some already known techniques of quantum metrology \cite{Aiello2004,Aiello2007,Kimani}.
Last but not least, our work furnishes a clear proof of principle that optical measurements requiring entanglement but not nonlocality may be accomplished by using classical light. 
%
%
%
\begin{appendix}
\section{Qubit operations}
\label{app:A}
\subsection{Single-DoF operations}\label{app:A1}
Consider the single-qubit two-dimensional Hilbert space $\mathscr{H}_1=\mathrm{span}\{\ket{0},\ket{1}\}$, where the standard basis states $\ket{0}$ and $\ket{1}$ are defined as the eigenstates of the $\sigma_3$ Pauli matrix in \eref{eq:pauli_matrices}, irrespective of the specific DoF encoding the qubit. All the results obtained in this appendix are indeed equally valid for both polarization and spatial qubits, as defined by \eref{eq:basisP} and \eref{eq:basisS}. Similarly, the basis vectors $\{\ket{+},\ket{-}\}$ and $\{\ket{L},\ket{R}\}$ are defined as the eigenstates of the remaining two Pauli matrices $\sigma_1$   and $\sigma_2$, respectively, where
\numparts
\BEA
\ket{+} &=  \frac{\ket 0 + \ket 1}{\sqrt{2}},  \qquad  &\ket{-}  =  \frac{\ket 0 - \ket 1}{\sqrt{2}}, \label{V.10A} \\
\ket{L} &=  \frac{\ket 0 + \rmi \ket 1}{\sqrt{2}},  \qquad  &\ket{R}  =  \frac{\ket 0 - \rmi \ket 1}{\sqrt{2}} . \label{V.10B}
\EEA
\endnumparts
Rotatable $\pi$- and $\pi/2$-converters permit any transformation between these basis vectors \cite{Paz}. According to \cite{PedrottiBook}, the unitary matrices representing  $\pi$- and $\pi/2$-converters can be written as:
\BEA
\label{V.20}
U_\pi =  e^{-\rmi \pi /2}  \left[\begin{array}{cc}
 1 & 0\\
 0 & -1
\end{array}\right], \qquad 
U_{\pi/2}  = e^{-\rmi \pi /4} \left[\begin{array}{cc}
 1 & 0\\
 0 & \rmi
\end{array}\right],
\EEA
where the conventional overall phase factors are fixed by the condition that, for the polarization qubit, the fast axis of both HWP and QWP are  horizontal. The unitary matrix $U_{\varphi}(\theta)$, with $\varphi \in \{\pi, \pi/2 \}$ for a $\varphi$-converter rotated by an angle $\theta$, is given by
\BEA
\label{V.30}
U_{\varphi}(\theta)= D\left( \theta \right) U_{\varphi} D\left( -\theta \right),
\EEA
where $D\left( \theta \right)$ denotes the standard $2 \times 2$ rotation matrix: 
\BEA
\label{V.40}
D\left( \theta \right) = \left[\begin{array}{cc}
 \cos \theta & - \sin \theta\\
 \sin \theta & \cos \theta
\end{array}\right].
\EEA
For example, we can use \eref{V.30} to transform the vectors \eref{V.10A} and \eref{V.10B} into the standard basis, as follows:
\begin{equation} \label{V.50}
\begin{array}{rcrcr}
U_{\pi}(\pi/8)  |+ \rangle &\!\!=\!\!& -\rmi |0 \rangle, \qquad U_{\pi}(\pi/8)  |- \rangle &\!\!=\!\!& -\rmi |1 \rangle,\\
U_{\pi/2}(\pi/4)  |L \rangle &\!\!=\!\!&  |0 \rangle, \qquad U_{\pi/2}(\pi/4) |R \rangle &\!\!=\!\!&-\rmi |1 \rangle.
\end{array}
\end{equation}

Consider now the four basis states $\{\ket{0}, \, \ket{1}, \, \ket{+}, \, \ket{L} \}$ that we conveniently relabel as $\{\ket{0}, \, \ket{1}, \, \ket{2}, \, \ket{3} \}$. From these states we can built the four linearly independent projection matrices $E_\mu \equiv \ket{\mu}\bra{\mu}$:
\BEA
\label{V.60} 
&E_0=
\left[\begin{array}{cc}
 1 & 0\\
 0 & 0
\end{array}\right] = \frac{\sigma_0+\sigma_3}{2},\qquad\qquad
&E_1=
\left[\begin{array}{cc}
 0 & 0\\
 0 & 1
\end{array}\right]= \frac{\sigma_0-\sigma_3}{2},\nonumber\\
&E_2=
\left[\begin{array}{cc}
 \frac{1}{2} & \frac{1}{2}\\
 \frac{1}{2} & \frac{1}{2}
\end{array}\right]= \frac{\sigma_0+\sigma_1}{2},\quad\quad
&E_3 =
\left[\begin{array}{cc}
 \frac{1}{2} & \frac{- \rmi}{2}\\
 \frac{\rmi}{2} & \frac{1}{2}
\end{array}\right]= \frac{\sigma_0+\sigma_2}{2}.
\EEA
These relations can be inverted to give $\sigma_0 = E_0 +E_1$, $\sigma_1 = -E_0 - E_1 + 2 E_2$, $\sigma_2 = - E_0 - E_1 + 2 E_3$, $\sigma_3 = E_0 - E_1$ or, formally, 
\BEA
\label{V.70}
E_\mu = \sum_{\alpha=0}^3 g_{\mu \alpha} \sigma_\alpha, \qquad \mathrm{and} \qquad\sigma_\mu = \sum_{\alpha=0}^3 f_{\mu \alpha} E_\alpha, \qquad (\mu =0,1,2,3),
\EEA
where, from the orthogonality of the Pauli matrices it follows that $g_{\mu \alpha} = \tr \left( E_\mu \sigma_\alpha \right)/2$. The coefficients $f_{\mu \alpha}$ can be found by noticing that the two $4 \times 4$ matrices $F$ and $G$ defined as $\left[ G \right]_{\mu \alpha} = g_{\mu \alpha}$ and $\left[ F \right]_{\mu \alpha} = f_{\mu \alpha}$, are connected by the simple relation  $F = G^{-1}$, which implies $f_{\mu \alpha} =  \left[ G^{-1} \right]_{\mu \alpha}$. 

From an operational point of view, the projector $E_2$ can be physically implemented with a $\pi$-converter followed by the projector $E_0$:
\BEA
\label{V.72}
E_2 = U_{\pi}^\dagger(\pi/8) \,E_0\, U_{\pi}(\pi/8) ,
\EEA
where \eref{V.50} has been used. Similarly, the projector $E_3$ can be realized with a $\pi/2$-converter followed by the projector $E_0$:
\BEA
\label{V.74}
E_3 = U_{\pi/2}^\dagger(\pi/4)\, E_0 \,U_{\pi/2}(\pi/4) .
\EEA

Finally, the single-DoF Stokes parameters $S_\mu = \tr \left(\rho_{d} \, \sigma_\mu \right)$, with $\rho_d$ denoting the single-DoF $2 \times 2$ coherency matrix and $d\in\{\mathrm{pol},\mathrm{spa}\}$, can be expressed in the basis $\{E_\alpha \}$ as
\BEA
\label{V.80}
S_\mu = \sum_{\alpha=0}^3 f_{\mu \alpha} I_\alpha, \qquad \mathrm{where} \qquad I_\alpha = \tr \left( \rho_d E_\alpha \right). 
\EEA
For example, for polarization qubits $I_\alpha$ denotes the intensity of light polarized in the state $|\alpha \rangle$.
\subsection{Two-DoF operations}\label{app:A2}
The mathematical apparatus developed in \ref{app:A1} can be used to express the two-DoF Stokes parameters $S_{\mu \nu}$ in terms of measurable intensities of light. To this end, it is enough to rewrite \eref{eq:stokes_correlation} in terms of the projection matrices $\{E_\mu\}$ as:
\BEA
\label{V.90}
S_{\mu\nu} & = &\tr \left[ \rho  \left(\sigma_\mu\otimes\sigma_\nu \right)\right] \nonumber \\
& = & \sum_{\alpha,\beta=0}^3 f_{\mu \alpha} f_{\nu \beta} I_{\alpha \beta} \nonumber \\
& = & \sum_{\alpha,\beta=0}^3 \left( F \otimes F\right)_{\mu \nu, \alpha \beta}  I_{\alpha \beta},
\EEA
where $I_{\alpha \beta} \equiv \tr \left[ \rho  \left(E_\alpha \otimes E_\beta \right)\right]$ denotes the intensity measured by the detector labeled by the pair of indexes $\alpha, \, \beta$ in figure \ref{fig.full_setup},  and \eref{V.60} has been used for both the polarization and the spatial qubits. The last row of \eref{V.90} furnishes a straightforward way to calculate the coefficients $ f_{\mu \alpha} f_{\nu \beta}$. However, a more efficient formula can be obtained by defining the ``intensity matrix'' $I$ via the relation $[I]_{\alpha \beta} = I_{\alpha \beta}$. Then, from  \eref{V.90} it follows that
\BEA
\label{V.100}
S_{\mu\nu} = \left[F I F^T \right]_{\mu\nu}.
\EEA
From \eref{V.100} one obtains, for example,  $S_{00} = I_{00} + I_{01} + I_{10} + I_{11}$ and $S_{31} = -I_{00} - I_{01} + 2I_{02} + I_{10}+ I_{11} - 2I_{12}$. The expression for $S_{00}$ originates directly  from the  relation $E_0 \otimes E_0 + E_0 \otimes E_1 + E_1 \otimes E_0 + E_1 \otimes E_1 = \mathcal{I}_4$, where $\mathcal{I}_4$ denotes the $4 \times 4$ identity matrix. A complete list of the two-Dof Stokes parameters $S_{\mu\nu}$ expressed in terms of the intensities $I_{\alpha \beta}$ is given below:
\BEA\label{V.110}
\begin{array}{lcl}
 S_{00} & = &  I_{00}+I_{01}+I_{10}+I_{11}, \\
 S_{01} & = &   -I_{00}-I_{01}+2 I_{02}-I_{10}-I_{11}+2 I_{12}, \\
 S_{02} & = &   -I_{00}-I_{01}+2 I_{03}-I_{10}-I_{11}+2 I_{13}, \\
 S_{03} & = &   I_{00}-I_{01}+I_{10}-I_{11}, \\
 S_{10} & = &   -I_{00}-I_{01}-I_{10}-I_{11}+2 \left(I_{20}+I_{21}\right), \\
 S_{11} & = &   I_{00}+I_{01}-2 I_{02}+I_{10}+I_{11}-2 \left(I_{12}+I_{20}+I_{21}-2 I_{22}\right), \\
 S_{12} & = &   I_{00}+I_{01}-2 I_{03}+I_{10}+I_{11}-2 \left(I_{13}+I_{20}+I_{21}-2 I_{23}\right), \\
 S_{13} & = &   -I_{00}+I_{01}-I_{10}+I_{11}+2 I_{20}-2 I_{21} ,\\
 S_{20} & = &   -I_{00}-I_{01}-I_{10}-I_{11}+2 \left(I_{30}+I_{31}\right), \\
 S_{21} & = &   I_{00}+I_{01}-2 I_{02}+I_{10}+I_{11}-2 \left(I_{12}+I_{30}+I_{31}-2 I_{32}\right), \\
 S_{22} & = &   I_{00}+I_{01}-2 I_{03}+I_{10}+I_{11}-2 \left(I_{13}+I_{30}+I_{31}-2 I_{33}\right), \\
 S_{23} & = &   -I_{00}+I_{01}-I_{10}+I_{11}+2 I_{30}-2 I_{31}, \\
 S_{30} & = &   I_{00}+I_{01}-I_{10}-I_{11} ,\\
 S_{31} & = &   -I_{00}-I_{01}+2 I_{02}+I_{10}+I_{11}-2 I_{12}, \\
 S_{32} & = &   -I_{00}-I_{01}+2 I_{03}+I_{10}+I_{11}-2 I_{13} ,\\
 S_{33} & = &   I_{00}-I_{01}-I_{10}+I_{11}.
\end{array}
\EEA
%

\section{Mode beam splitter}
\label{app:ms}
We consider the experimental realization of a mode beam splitter (MBS) displayed in Fig. \ref{fig.MCP_MBS} {d). To this end we use a right-handed coordinate system attached to the beam whose direction of propagation always coincides with the $z$-axis}. 
%
%
%
\begin{figure}[!ht]
\begin{center}
\includegraphics[width=10.3truecm]{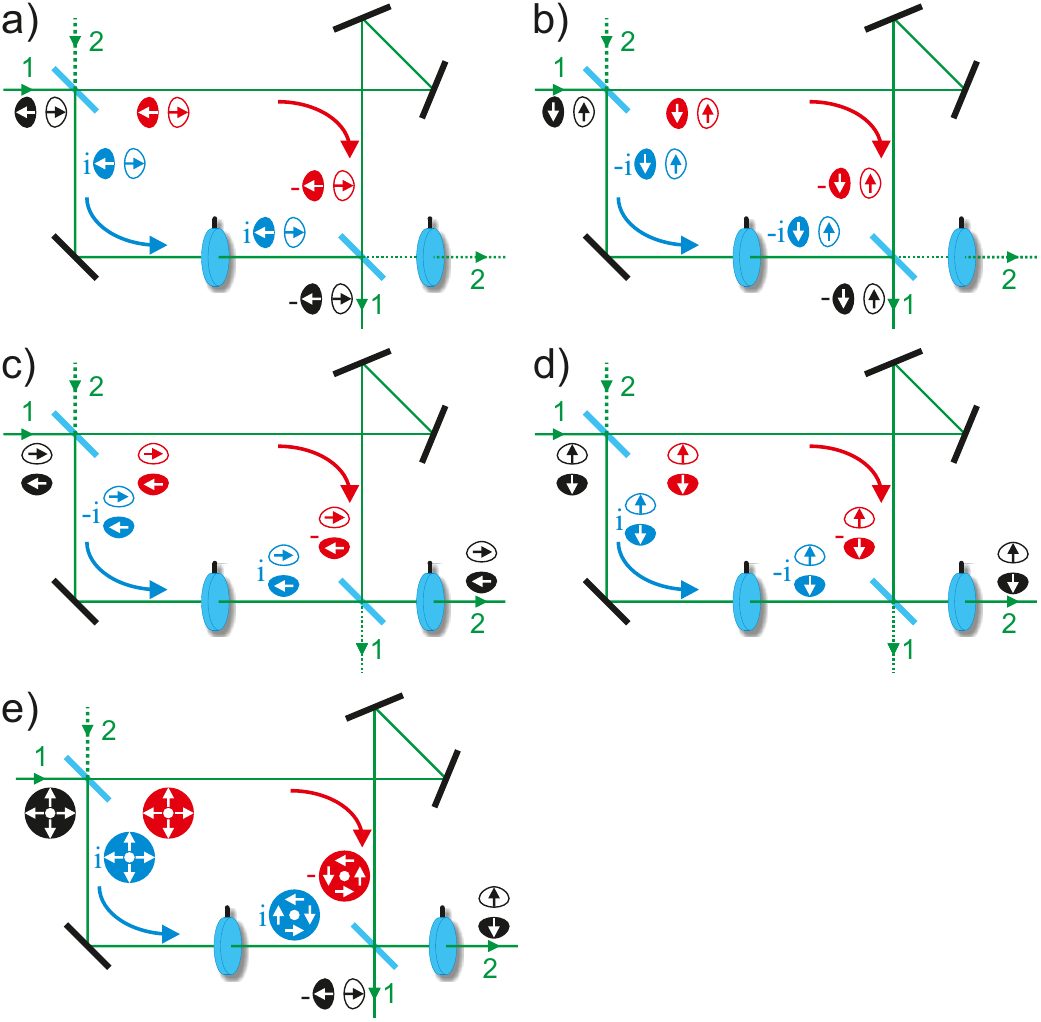}
\caption{\label{fig.mode_splitter} Visualization of the working principle of a mode beam splitter (MBS). The numbers 1 and 2 label the two arms of the MBS and dashed lines denote dark channels. a-b) The horizontally and vertically polarized HG modes $\bE_1^\mathrm{in}(\brr)=\bhe_x\psi_{10}(\brr)$ and $\bE_1^\mathrm{in}(\brr)=\bhe_y\psi_{10}(\brr)$ are transmitted across port 1. c-d) The horizontally and vertically polarized HG modes $\bE_1^\mathrm{in}(\brr)=\bhe_x\psi_{01}(\brr)$ and $\bE_1^\mathrm{in}(\brr)=\bhe_y\psi_{01}(\brr)$ are transmitted across port 2. e) The MBS splits a radially polarized beam $\bE_1^\mathrm{in}(\brr)=[\bhe_x\psi_{10}(\brr)+\bhe_y\psi_{01}(\brr)]/\sqrt{2}$  in its components $\bE_1^\mathrm{out}(\brr)=\bhe_x\psi_{10}(\brr)$ and $\bE_2^\mathrm{out}(\brr)=\bhe_y\psi_{01}(\brr)$.}
\end{center}
\end{figure}
%
%
%
On each reflection the handedness of the spatial modes is inverted and the phase difference between $x$ and $y$-polarization components is shifted by $\pi$. {This means} that each reflection maps the coordinate $x$ onto $-x$ and the polarization vector $\bhe_x$ onto $-\bhe_x$. Hence, a mirror is described by the following transformation of the Jones vector:
\BEA
\left[\begin{array}{c}
E_x(\brr)\\
E_y(\brr)
\end{array}\right]
\rightarrow\rmi
\left[\begin{array}{c}
-E_x(\bar{\brr})\\
E_y(\bar{\brr})
\end{array}\right],
\EEA
where $\bar{\brr}= -x \bhe_x +y \bhe_y + z \bhe_z$. Accordingly, a symmetric 50/50 beam splitter acts on the input field $[\bE_1(\brr),\bE_2(\brr)]^T$ as follows:
\BEA
\left[\begin{array}{c}
E_{1x}(\brr)\\
E_{1y}(\brr)\\
E_{2x}(\brr)\\
E_{2y}(\brr)
\end{array}\right]
\rightarrow\frac{1}{\sqrt{2}}
\left[\begin{array}{c}
E_{1x}(\brr)-\rmi E_{2x}(\bar{\brr})\\
E_{1y}(\brr)+\rmi E_{2y}(\bar{\brr})\\
E_{2x}(\brr)-\rmi E_{1x}(\bar{\brr})\\
E_{2y}(\brr)+\rmi E_{1y}(\bar{\brr})
\end{array}\right],
\EEA
{where the subscripts 1 and 2 denote the two ports of the BS and, e. g., $\bE_1(\brr)$ denotes the electric field of the beam entering port $1$.} The HWP with its fast optical axis aligned parallel to the horizontal direction is, according to \eref{V.20}, described by the transformation
\BEA
\left[\begin{array}{c}
E_x(\brr)\\
E_y(\brr)
\end{array}\right]
\rightarrow\rmi
\left[\begin{array}{c}
-E_x(\brr)\\
E_y(\brr)
\end{array}\right].
\EEA
{As our proposal uses solely first-order spatial modes, 
let us consider the input fields $\bE_1^\mathrm{in}(\brr)=[A_{00}\psi_{10}(\brr)+A_{01}\psi_{01}(\brr)]\,\bhe_x+[A_{10}\psi_{10}(\brr)+A_{11}\psi_{01}(\brr)]\,\bhe_y$ and $\bE_2^\mathrm{in}(\brr)=\boldsymbol{0}$. The MBS transforms these input fields into the output fields 
\BEA
\bE_1^\mathrm{out}(\brr)=-(A_{00}\bhe_x+A_{10}\bhe_y)\psi_{10}(\brr),
\EEA
and 
\BEA
\bE_2^\mathrm{out}(\brr)=(A_{01}\bhe_x+A_{11}\bhe_y)\psi_{01}(\brr).
\EEA
 as can be shown by successively applying the transformations of each element of the MBS described above. Furthermore, it was used the fact that the HG mode $\psi_{10}$ changes sign upon reflection, i.e. $\psi_{10}(\bar{\brr})=-\psi_{10}(\brr)$, whereas this is not the case for the HG mode $\psi_{01}$, i.e. $\psi_{01}(\bar{\brr})=\psi_{01}(\brr)$.}
 
\end{appendix}

\newpage

%

%
\end{document}